# Suppression of Thin-Film Thermal Conductivity due to Surface Roughness

Michimasa Morita[1] and Junichiro Shiomi[1,2,*]

[1]Department of Mechanical Engineering, The University of Tokyo, Tokyo 113-8656, Japan
[2]Institute of Engineering Innovation, The University of Tokyo, Tokyo 113-0033, Japan
[*]Contact author: shiomi@photon.t.u-tokyo.ac.jp

**ABSTRACT**. Understanding thermal transport in silicon nanostructures is crucial for effective thermal management in semiconductor devices. In such nanostructures, boundary scattering can significantly reduce thermal conductivity. Diffusive boundary scattering explains the experimentally observed thickness dependence of thermal conductivity in thin films with thicknesses of tens of nanometers; however, introducing surface roughness further reduces the thermal conductivity, which falls far below the theoretical lower limit. In this study, we calculated the thermal conductivity and phonon transport properties of rough thin films with thicknesses of up to 25 nm using anharmonic lattice dynamics and investigated the mechanisms underlying the suppression of thermal conductivity arising from surface roughness. We found that in ultrathin films with rough surfaces, thermal conductivity was suppressed by a reduction in group velocity caused by hybridization with surface-localized modes, as well as a reduction in relaxation time due to the modulation of the anharmonic interatomic force constants of surface atoms. The reduction in group velocity significantly suppressed thermal conductivity across a wide range of thicknesses and surface-roughness values. In contrast, the reduction in relaxation time exhibited strong thickness dependence. Thus, this relaxation-time reduction should be considered in ultrathin films with roughness of approximately 0.1 nm and thicknesses below 5 nm. These thermal-conductivity suppression mechanisms due to surface roughness were not considered in the boundary-scattering model, resulting in an overestimation of the thermal conductivity of the roughened thin films by up to approximately 100%.

## I. INTRODUCTION

Miniaturization of transistors in integrated circuits is essential for enhancing computer performance. However, in recent years, thermal dissipation has become a bottleneck for continued transistor miniaturization [1, 2]. Although transistors with characteristic lengths of <10 nm have been developed [3], their heat-dissipation capability is significantly limited by size effects [4]. For example, the thermal conductivity of ultrathin silicon films [5–9], which is frequently measured to demonstrate size effects, exhibits a significant reduction from the bulk value of 150 W/mK [10]. Specifically, for films with thicknesses of 20–30 nm, the thermal conductivity decreases to approximately 30 W/mK [7, 8, 11], and for films with thicknesses of approximately 10 nm, it decreases to approximately 10 W/mK [9]. This reduction in thermal conductivity can be detrimental to heat dissipation in integrated circuits [12].

Phonon boundary scattering is widely accepted as a key mechanism responsible for size effects [6]. Fuchs [13] and Sondheimer [14] solved the Boltzmann transport equation in thin films with boundary conditions defined by the surface specularity parameter $p$ and proposed thickness-dependent transport suppression. This model successfully reproduces the thermal conductivities of thin films with thicknesses of >20 nm within the diffuse-scattering limit ($p=0$) [11, 15]. Although $p$ is typically determined via fitting, Ziman [16] and Maznev [17] suggested that it can be determined using the root-mean-square roughness (RMSR) $\eta$ of the film surface. In this model, an increase in surface roughness reduces the specularity, leading to a decrease in thermal conductivity. Therefore, the diffusive-scattering limit corresponded to a sufficiently large RMSR relative to the phonon wavelength.

Attempts to experimentally correlate the surface roughness with the reduction in thermal conductivity due to size effects have been widely conducted. For example, Hochbaum et al. [18] fabricated nanowires with diameters of 50–115 nm that had artificially rough surfaces via etching. They reported that the thermal conductivity was reduced by approximately one order of magnitude owing to surface roughness. Similar reductions in thermal conductivity due to surface roughness have been observed in other nanowires [19–23] and thin films [24, 25]. However, the experimental values obtained with the introduced roughness fell below the diffuse-scattering limit [18, 24] and could not be described by the Fuchs–Sondheimer model.

Moreover, Jiang et al. [26] pointed out that even for measurement results that can be well-reproduced by the diffusive-scattering limit in the Fuchs–Sondheimer model [6, 11, 27], the estimated $\eta$ from transmission electron microscopy (TEM) images [26, 28] was approximately 0.1 nm. When this value was substituted into Ziman's specularity formula, it overestimated $p$, yielding a predicted thermal conductivity that significantly exceeded the experimental result.

One possible reason for the overestimation of the thermal conductivity in the Fuchs–Sondheimer model based on Ziman's formula is the modulation of the phonon dispersion relations. Because the Fuchs–Sondheimer model uses bulk phonon properties, it fails to account for the reduction in group velocity due to dispersion relation modulations. Cuffe et al. [29] measured the group velocities of the acoustic modes in freestanding silicon films with a thickness of approximately 8 nm. They reported a reduction in group velocity by approximately one order of magnitude, indicating that the reduced group velocity cannot be ignored at extremely small scales. Similar reductions in group velocity have been observed in computational simulations of thin films [30, 31] and nanowires [32]. Another possible factor is anharmonic phonon scattering. Ziman's specularity approach addresses the boundary scattering based solely on harmonic approximations. However, previous research [33] on the thermal conductivity of smooth, freestanding silicon films indicated that anharmonic scattering by surface-localized phonons contributes to the reduction in thermal conductivity. However, these previous studies either did not introduce surface roughness [29–30, 32, 33] or considered scales that were too small compared with the experimental values [31]. Therefore, it remains unclear whether the modulation of harmonic and anharmonic phonon properties reduces the thermal conductivity of thin films with rough surfaces.

In this study, we calculated the harmonic and anharmonic phonon transport properties in rough thin films using the anharmonic lattice dynamics (ALD) method and investigated the dependence of these modulations on the surface roughness and film thickness. Furthermore, by examining the modulation of the phonon transport properties and thermal conductivity, we sought to uncover the underlying mechanisms of thermal-conductivity suppression in ultrasmall-scale rough thin films.

## II. METHODS

### A. Thin-film structure model

Initially, a unit cell of the thin film was created by repeating the face-centered cubic unit cell of silicon in the $x$–$y$ direction for two periods and in the $z$ direction for $N_{layer}$ periods. A 1-nm-thick vacuum layer was inserted above and below the structure to create smooth surfaces. Surface roughness was then introduced by randomly removing the surface atoms from the smooth thin film and adding them to the opposite surface. To increase the roughness through this reconfiguration, the atoms were removed from the top surface in the red regions and from the bottom surface in the blue regions, as shown in Fig. 1, with $N$ atoms removed in each case. The rough structures were relaxed to obtain stable structures. Four repositioned surface patterns were prepared. By varying $N_{layer}$ for each pattern, thin films of various thicknesses were created. The surface structures and their RMSRs are shown in Fig. 2.

Patterns 1–3 were designed with relatively high roughness. Their RMSRs ($\eta$) were approximately 0.07, 0.14, and 0.12 nm, respectively. Pattern 4, designed with very low roughness, exhibited $\eta \leq 0.01$ nm. As TEM measurements have indicated $\eta$ values of 0.11 nm [26] at the Si/SiO$_2$ interfaces of non-freestanding thin films, the surface roughness of Patterns 1–3 was reasonable. The variation in RMSR with respect to film thickness for Pattern 1 was due to the breakdown of the rough structure during relaxation.

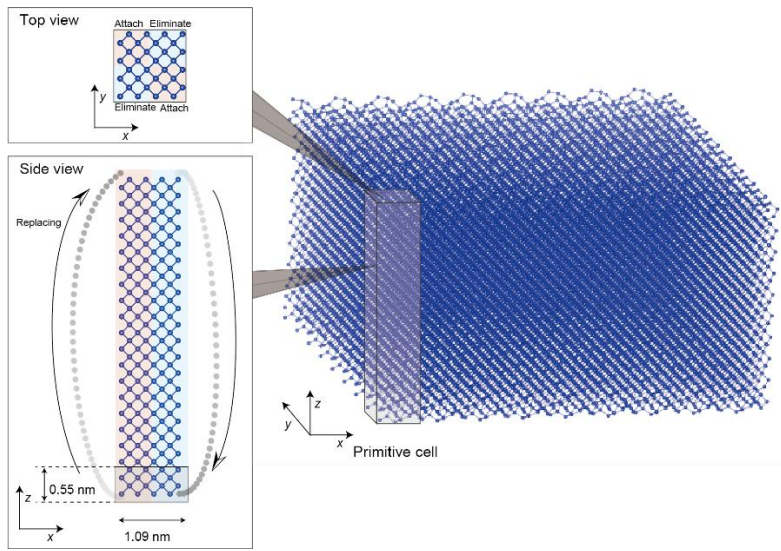

FIG. 1. Schematic of the generated rough thin film.

## B. ALD calculations

For the thermal-conductivity calculations, we used the ALD calculation package ALAMODE [34]. Harmonic and anharmonic interatomic force constants (IFCs) required for ALD calculations are often obtained from density functional theory (DFT) calculations [31, 35–37]. However, because such calculations are computationally intensive, this approach was impractical for the systems investigated in the present study, which involved over 1000 atoms. Therefore, we calculated the IFCs using the Stillinger–Weber potential [38], which is known to reproduce the material properties of silicon, with fitted parameters to match DFT calculations [39]. The IFCs were calculated using a direct method [40].

To reduce the computational cost, we applied an interatomic cutoff, considering harmonic IFCs up to the eighth-nearest neighbors and cubic IFCs up to the second-nearest neighbors. For large systems, relaxation-time calculations, which are the most time-consuming part of ALD calculations, typically require $O(N^4_{atom})$ computational time [41]. Therefore, a simple implementation of ALD is impractical for large systems. Thus, in this study, we implemented Monte Carlo integration with importance sampling [42–44] for relaxation-time calculations and reduced the computational cost to $O(N^2_{atom}N_{sample})$. The scattering phase space (SPS), which can be calculated using only harmonic terms, was used to generate the probability distributions for importance sampling. The delta function in the relaxation-time integration was approximated by a Lorentzian function, with a smearing width set to 0.1 cm$^{-1}$ for bulk silicon and adjusted for thin films according to the density of integration points in the reciprocal space. The reciprocal mesh size was set to 7×7×1.

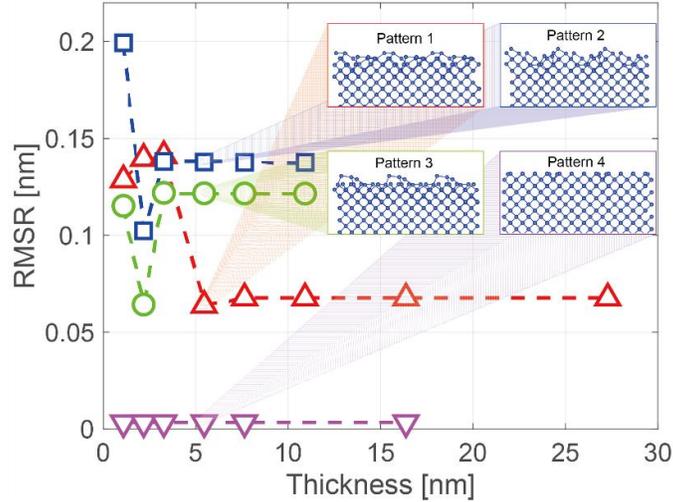

FIG. 2. RMSRs of the generated thin-film surfaces. Red, blue, green, and purple represent the surfaces of Patterns 1–4, respectively. The insets show images of the 5.5-nm-thick film for each pattern.

## III. RESULTS AND DISCUSSION

### A. Thermal conductivity

The thermal conductivities obtained are presented in Fig. 3. As shown in Fig. 3(a), the thermal conductivities of all thin films in Patterns 1–4 were significantly lower than the bulk thermal conductivity. Specifically, while the thermal conductivity of the completely smooth thin film was 50%–80% of the bulk thermal conductivity, the thermal conductivities of the rough thin films (Patterns 1–4) were <50% of the bulk conductivity. This suggested that the introduction of roughness suppresses the thermal conductivity of the thin films. In addition, the thermal conductivity of thin films with $\eta \simeq 0.01$ nm (Pattern 4) was approximately 30%–50% of the bulk value. In contrast, the thermal conductivities of rough thin films with $\eta \simeq 0.1$ nm (Patterns 1–3) were approximately 1%–20% of the bulk value. This indicated that an increase in roughness further reduces the thermal conductivity. However, despite differences in $\eta$ (=0.05–0.2) nm among Patterns 1–3 (Fig. 2), there was little difference in their thermal conductivities. This suggested that the reduction in thermal conductivity due to increased roughness became saturated at $\eta \approx 0.1$ nm.

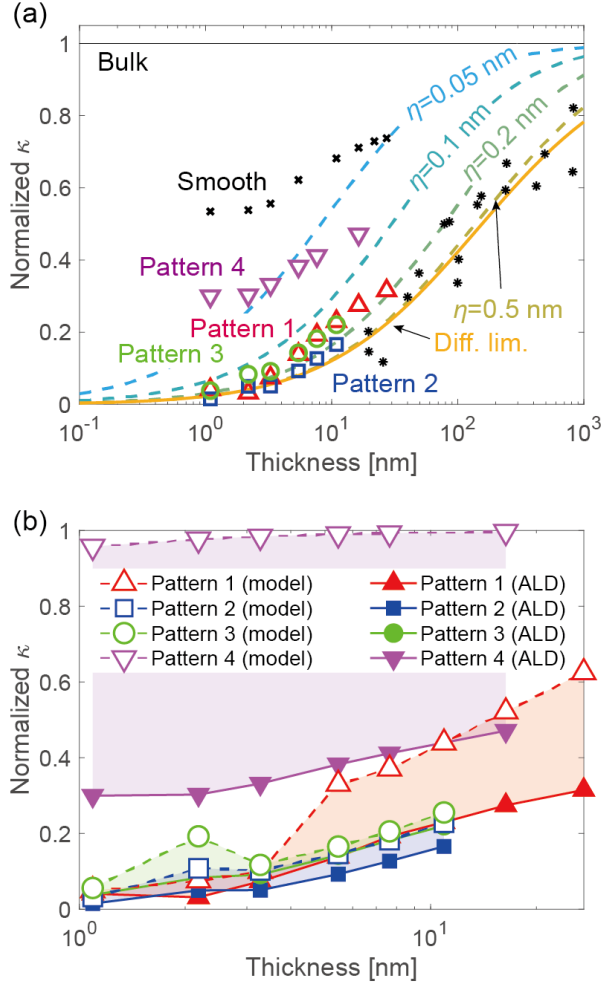

FIG. 3. (a) Thickness dependence of thermal conductivity normalized by bulk thermal conductivity at $T$=300 K. (b) Comparison of the thermal conductivity of each thin film with the theoretical model. The empty markers represent the predicted values from the Fuchs–Sondheimer model, calculated using the RMSR of each thin film (Fig. 2). The solid markers represent the thermal conductivities calculated using ALD.

Next, the thermal conductivities calculated using the theoretical model were compared. According to the Fuchs–Sondheimer model [13, 14], the thickness-dependent thermal conductivity is expressed as

$$\kappa = \sum_{q,s} \kappa_{\text{bulk}}(qs)\left[1 - (1-p)\frac{3\Lambda_{q,s}}{2t}F(\Lambda_{q,s}, p)\right], \quad (1)$$

where $F(\Lambda_{qs}, p)$ is referred to as the Fuchs–Sondheimer correction factor:

$$F(\Lambda_{q,s}, p) = \int_0^{\frac{\pi}{2}} d\theta \, \sin^3(\theta) \cos(\theta) \left\{ \frac{1-\exp(-t/\Lambda_{q,s}\cos(\theta))}{1-p\exp(-t/\Lambda_{q,s}\cos(\theta))} \right\}. \quad (2)$$

Here, the specularity parameter $p$ is generally treated as a fitting parameter. However, Ziman derived $p$ as a function of $\eta$ and the phonon wavelength $\lambda_{qs}$. According to Ziman's derivation [16, 17], the specularity $p$ is given by

$$p = \exp\left(-\frac{16\eta^2\pi^2}{\lambda_{q,s}^2}\right). \quad (3)$$

The thermal conductivities of Patterns 1–3 were close to the diffusive limit of the Fuchs–Sondheimer model, which is similar to the experimental values shown in Fig. 3(a). In contrast, Pattern 4 and the smooth film exhibited different thickness dependence from the

Fuchs–Sondheimer model. For example, the thermal conductivity of Pattern 4 was slightly higher than that of the Fuchs–Sondheimer model with $\eta=0.05$ nm at a thickness of $t=1$ nm. However, at $t=15$ nm, the thermal conductivity was significantly lower than that of the Fuchs–Sondheimer model with the same $\eta$. Additionally, the RMSR of the thin films in Pattern 4 was ~0.01 nm, which was significantly smaller than 0.05 nm.

We compared the thermal conductivity predicted by the Fuchs–Sondheimer model based on the $\eta$ of each film with the thermal conductivity calculated using ALD. As shown in Fig. 3(b), although the same $\eta$ was used, the thermal conductivities predicted by the Fuchs–Sondheimer model were generally higher than those calculated via ALD. This trend was particularly evident for Pattern 4; however, overestimation by a factor of approximately two was also observed for Pattern 1, and overestimation of ~50% was observed for Pattern 2. This suggests a limitation of the Fuchs–Sondheimer model based on Ziman's discussion of specularity.

Next, we compared the spectral thermal conductivities. Fig. 4 shows the calculated spectral thermal conductivity of the Pattern 1 thin film and that predicted using the Fuchs–Sondheimer model. In Fig. 4(a), a peak appears in the 1–2 THz range for thicknesses below 5 nm, and an additional peak at approximately 5 THz emerges for thicknesses above 5 nm. In contrast, the frequency dependence remained nearly unchanged, as shown in Fig. 4(b). In addition, Fig. 4(a) exhibits reduced thermal conductivity at >2 THz compared to Fig. 4(b), with a particularly significant reduction between 2 and 4 THz. This indicates that the theoretical model not only overestimated the overall thermal conductivity but also failed to capture its frequency dependence.

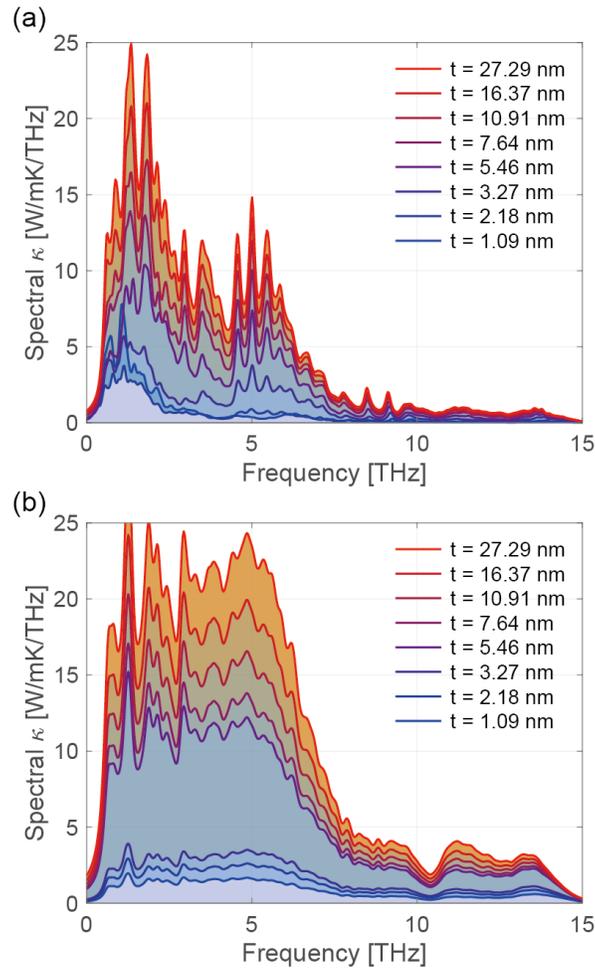

FIG. 4. Thickness dependence of the thermal-conductivity spectrum of the thin film with Pattern 1 at $T=300$ K. (a) Calculated spectral thermal conductivity. (b) Thermal-conductivity spectrum predicted by the Fuchs–Sondheimer model.

## B. Harmonic phonon properties

### *1. Phonon dispersion*

To consider the overestimation of the thermal conductivity in the Fuchs–Sondheimer model based on Ziman's specularity, we first examined the harmonic phonon properties. The dispersion relationships of the thin films are shown in Fig. 5. In Pattern 1, most of the optical modes above 2 THz were flattened, and few modes exhibiting high group velocities (indicated in blue) were observed. In contrast, for Pattern 4, many optical modes, particularly those in the 5–12 THz range, exhibited high group velocities. These high-group-velocity modes increased further for the smooth thin film. Because the thermal conductivity depends on the square of the group velocity, the reduced thermal conductivity for Pattern 1 is attributed to reduced group velocity.

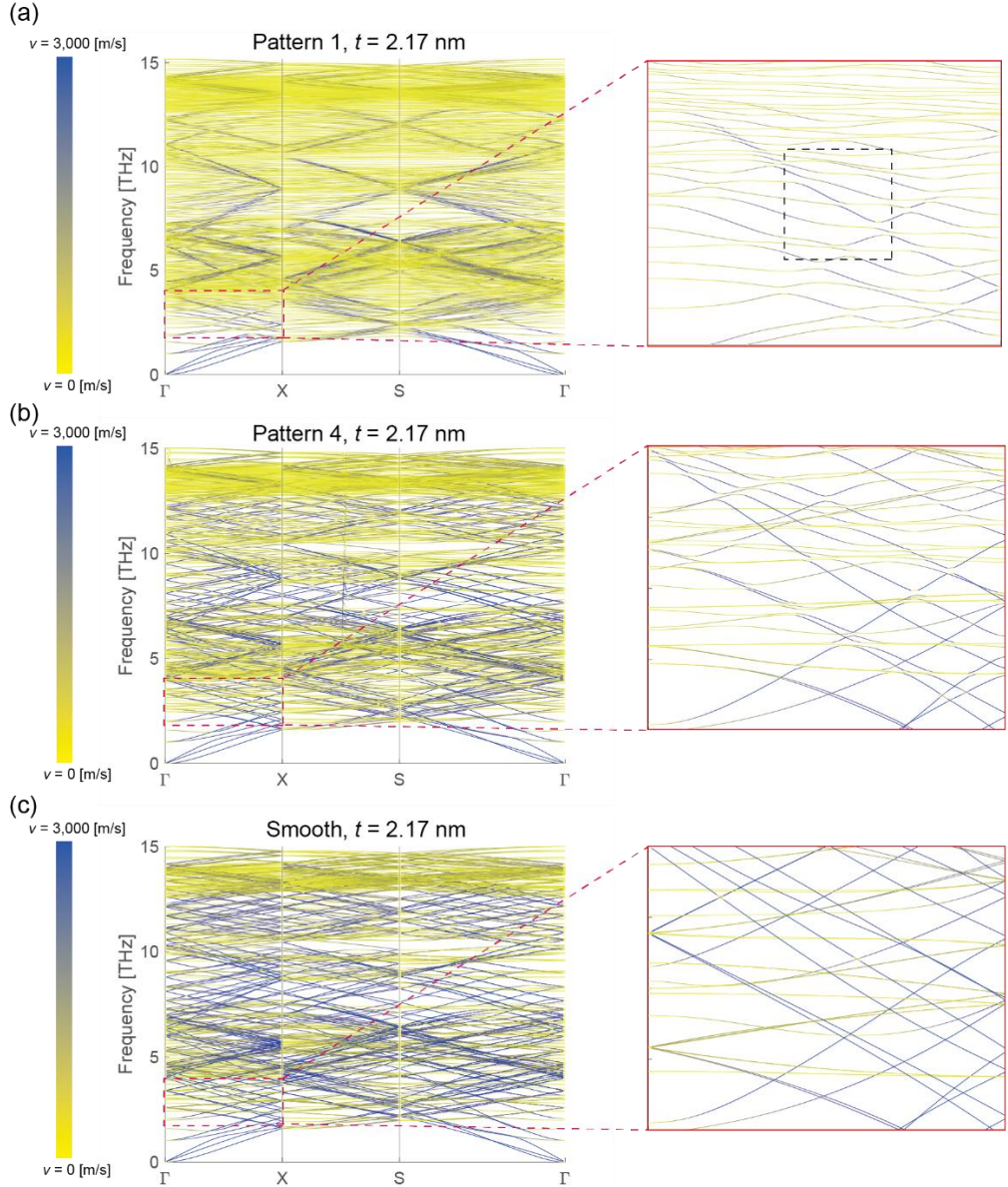

FIG. 5. (a–c) Dispersion relations of thin films with $t$=2.19 nm. Group velocities of different modes are projected onto the dispersion. The insets show enlarged views of the sections marked by the red line. (a) Pattern 1; (b) Pattern 4; (c) smooth thin film.

We then examined the mechanism underlying the suppression of group velocity. As shown in the inset of Fig. 5(a), the high-group-velocity modes interfered with the neighboring modes, leading to anti-crossing and a reduction in the group velocity. In contrast, as shown in the inset of Fig. 5(b), some of these modes crossed without interference from other modes, maintaining high group velocities. Moreover, in the inset of Fig. 5(c), no interference is observed, and most of the optical modes maintain a high group velocity across the entire Brillouin zone. Therefore, the differences in optical phonon group velocities originate from this interference.

Honarvar et al. [45] reported a similar phenomenon. They found that nanopillars placed on a silicon surface led to a similar interference and group-velocity reduction. They attributed this phenomenon to the hybridization between vibrons localized in the nanopillars and heat-conducting phonons. Therefore, in the rough thin films that we studied, heat-conducting phonons may hybridize with other modes. To explore this, we calculated the eigenmodes around the crossing regions where the interference occurred. The results are shown in Fig. 6. In the high-frequency mode indicated by the red circle, the eigenmode on the Γ-point side (left side) was uniformly distributed throughout the film, while the eigenmode on the X-point side (right side) was mostly localized at the surface atoms. In contrast, in the low-frequency mode indicated by the green circle, the eigenmode was localized at the surface on the Γ-point side and uniformly distributed throughout the film on the X-point side. Therefore, the eigenmodes of the high- and low-frequency modes switched in the interference region. This indicates that the localized phonon mode on the surface and uniformly distributed phonon mode hybridize in the interference region. Hence, just as the vibrons localized in the nanopillars reduced the group velocity, as reported by Honarvar et al., surface-localized phonons on rough surfaces played a similar role in reducing the group velocity of thin films with rough surfaces.

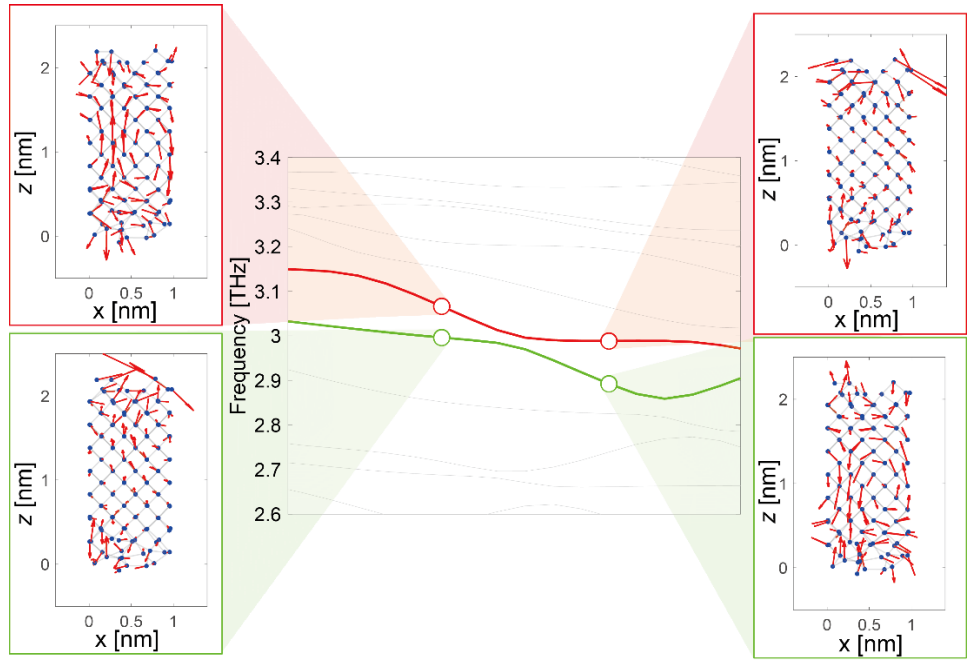

FIG. 6. Eigenmodes near the interference region. The frequency and wavevector region corresponds to the section marked by the black dotted line in the Fig. 5(a) inset.

### *2. Group-velocity reduction*

To quantitatively compare the group-velocity reductions due to roughness, we calculated the frequency-averaged group velocity for each surface structure. The results are shown in Fig. 7. In Fig. 7(a), for Patterns 1–3, significant reduction in group velocity is observed in the frequency range of >2 THz, where optical phonons dominate. This highlights the strong impact of roughness on the reduction in the optical phonon group velocity and explains the trend observed in the thermal-conductivity spectrum. As mentioned previously, compared with the theoretical model results shown in Fig. 4(b), the calculation results in Fig. 4(a) exhibit lower thermal-conductivity contributions above 2 THz, with a significant reduction in the 2–4 THz range. Similarly, the group velocity in the thin film of Pattern 1 decreased above 2 THz, particularly in the 2–4 THz frequency range, as shown in Figs. 7(a) and (b). Because the theoretical model did not account for the reduction in group velocity due to the modulation of the dispersion relation, the group-velocity reduction caused by surface roughness contributed to the discrepancy between the theoretical model and the calculated thermal-conductivity spectrum.

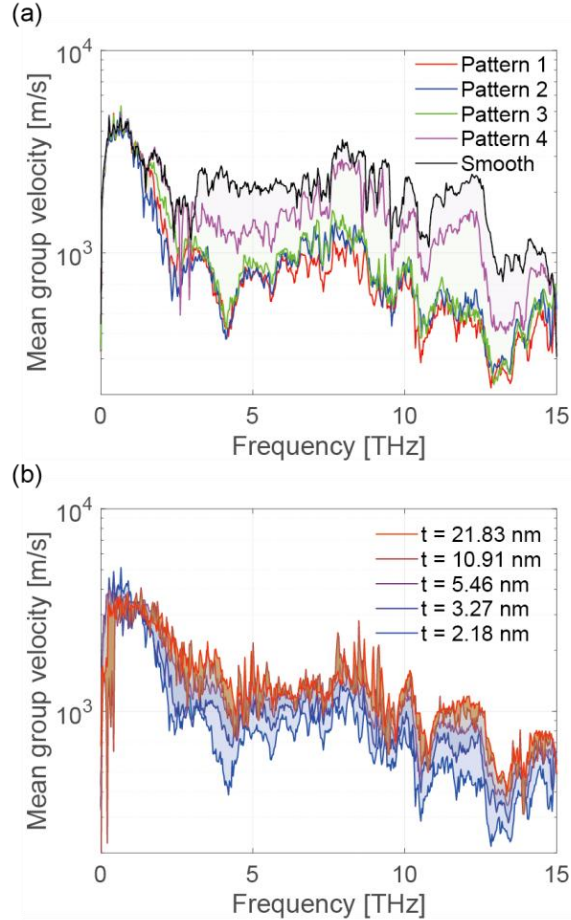

FIG. 7. (a, b) Frequency-averaged group velocities. (a) Surface-structure dependence at $t$=2.18 nm. (b) Thickness dependence for Pattern 1.

In contrast, in the low-frequency region below 2 THz, where acoustic phonons dominated, no surface-structure dependence on the group velocity was observed. This trend was consistent with the dispersion relations, where the acoustic phonons exhibited similar dispersion regardless of the surface structure (Figs. 5(a)–(c)). According to the dispersion-relation analysis, the group-velocity reduction was caused by hybridization with surface-localized phonons, which were optical phonons distributed at frequencies above 5 THz (see Supplementary Materials). Therefore, low-frequency acoustic phonons cannot hybridize with surface-localized phonons, and their group velocities remain unaffected. Next, we examined the thickness dependence of the group-velocity reduction due to roughness. As shown in Fig. 7(b), the group velocity of the phonons above 5 THz increased with the thickness. This was caused by the increased density of bands in the reciprocal space with thicker films, which weakened the impact of anti-crossing caused by interference (see Supplementary Materials). However, the thickness dependence diminished at lower frequencies, and an inverse thickness dependence was observed below 1

THz. This is attributed to the occurrence of zone folding with increased thickness, which reduced the lower limit of the optical phonon frequency, thereby reducing the group velocity of acoustic phonons (see Supplementary Materials).

Next, we calculated the effective group velocity $v_{\text{eff}}$, which is defined as follows:

$$\bar{v}_{\text{eff}}^2 = \frac{\sum_{q,s} C_v(qs)|v_{qs}|^2 \tau_{qs}}{\sum_{q,s} C_v(qs)\tau_{qs}}. \quad (4)$$

The calculated $v_{\text{eff}}$ values exhibited a strong dependence on the surface structure (Fig. 8). For the smooth surface, $v_{\text{eff}}$ was ~2700 m/s, whereas it was ~2200 m/s for Pattern 4 and 1000–2000 m/s for Patterns 1–3. Despite the low $\eta$ of <0.01 nm for Pattern 4, the $v_{\text{eff}}$ was intermediate between those of the smooth surface ($\eta=0$) and Patterns 1–3 ($\eta\simeq 0.1$ nm). However, despite the different surface structures and $\eta$ values for Patterns 1–3, the group-velocity reduction became saturated at $\eta\simeq 0.1$ nm. As mentioned previously, the interference at the crossing points of the modes reduces the group velocity. In Pattern 4, interference was observed at some crossing points, whereas in Pattern 1, interference occurred at all crossing points. This suggests that the number of crossing points with interference is saturated in Pattern 1; thus, the group-velocity reduction is saturated.

The surface-structure dependence of the thermal conductivity and group velocity exhibited similar trends. The highest values were observed for the smooth surface, the lowest values were observed for the rougher surfaces (Patterns 1–3), and intermediate values were observed for the less rough surface (Pattern 4). Because the lattice thermal conductivity is proportional to the square of the group velocity, the roughness dependence of the reduction in thermal conductivity is attributed to the suppression of the group velocity. However, the thickness dependence of the thermal conductivity and group velocity exhibited significantly different trends.

For rough thin films, the thermal conductivity of the thin film with $t=1.09$ nm was extremely low, at approximately 3% of the bulk value, whereas for $t=10.9$ nm, it increased to 15%–25% of the bulk, representing a 5–10-fold increase. In contrast, the $v_{\text{eff}}$ of the $t=1.09$ nm thin film was ~1,300 m/s, while that of the $t=10.9$ nm thin film was ~1,800 m/s, corresponding to a 1.5-fold increase. Considering the relationship between thermal conductivity and group velocity, the observed thickness dependence of the group velocity was insufficient to explain the thermal conductivity. Therefore, an additional mechanism for thermal-conductivity suppression with stronger thickness dependence was expected beyond the weak dependence of group-velocity-related suppression.

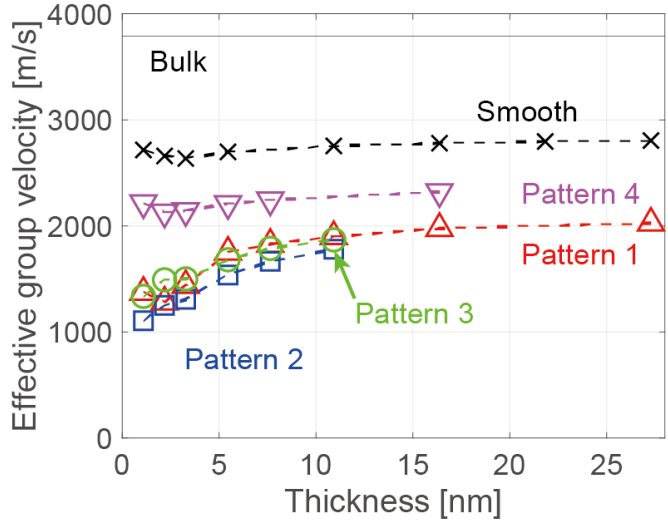

FIG. 8. Thickness dependence of $v_{\text{eff}}$. Red, blue, green, purple, and black dashed lines denote the $v_{\text{eff}}$ for Patterns 1–4 and the smooth thin film, respectively.

### C. Anharmonic phonon properties

#### 1. Phonon relaxation time

To explore another mechanism, we focused on phonon anharmonicity. For lattice thermal conductivity, the relaxation time indicates the anharmonicity of the phonon property. Therefore, we compared the thickness dependence of the relaxation times for each surface structure. Fig. 9 shows the thickness dependence of the frequency-averaged relaxation times. In Fig. 9(a), for the thin films of Pattern 1, the relaxation time increases with thickness across all frequencies. This trend is similar to the thickness dependence of thermal conductivity. Notably, while the group velocity for the low-frequency modes exhibited no thickness dependence, the relaxation time for these modes exhibited strong thickness dependence. Because low-frequency acoustic modes contribute significantly to the thermal conductivity (Fig. 4(a)), this thickness dependence of the relaxation time leads to strong thickness dependence of the thermal conductivity. The relaxation time also exhibited thickness dependence for Pattern 4, as shown in Fig. 9(b); however, this dependence was weaker than that shown in Fig. 9(a). This weaker thickness dependence relative to Pattern 1 is consistent with the thermal conductivity.

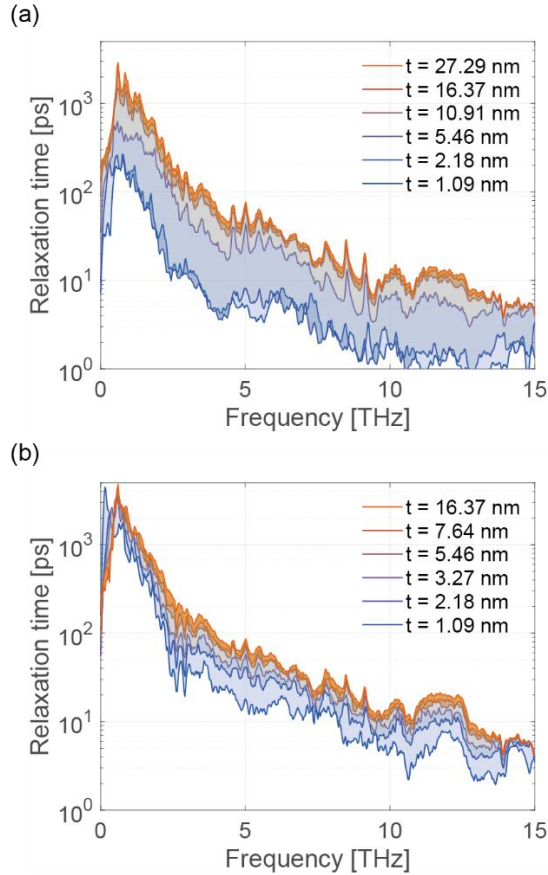

FIG. 9. (a) Thickness dependence of frequency-averaged relaxation times for the thin films of Pattern 1 at $T$=300 K. (b) Thickness dependence of frequency-averaged relaxation times for the thin films of Pattern 4 at $T$=300 K.

Next, to quantitatively evaluate the thickness dependence of the relaxation time due to surface roughness, we calculated the effective relaxation time $\tau_{\text{eff}}$. It is defined as follows:

$$\bar{\tau}_{\text{eff}} = \frac{\sum_{q,s} C_v(qs)|v_{qs}|^2 \tau_{qs}}{\sum_{q,s} C_v(qs)|v_{qs}|^2}. \quad (5)$$

As shown in Fig. 10, $\tau_{\text{eff}}$ for the film with a smooth surface (Smooth) and the film with $\eta \leq 0.01$ nm (Pattern 4) remained close to the bulk value and exhibited weak thickness dependence. In contrast, the thin films with $\eta \simeq 0.1$ nm (Patterns 1–3) exhibited strong thickness dependence. The thickness dependence for the films with different $\eta$ was similar to that of the thermal conductivity. Therefore, the thickness-dependent suppression of thermal conductivity is primarily driven by the relaxation time.

For Patterns 1–3 with a thickness of >10 nm, as well as for Pattern 4 and the smooth surface, the effective relaxation time remained close to the bulk value. In contrast, for Patterns 1–3 with a thickness of <5 nm, the effective relaxation time decreased to ≤50% of those of the bulk and smooth films. Therefore, surface roughness has little impact on the relaxation time for thicker films (>10 nm). However, for ultrathin films with thicknesses <5 nm and $\eta \simeq 0.1$ nm, surface roughness reduced the relaxation time to ≤50% of that for the smooth film.

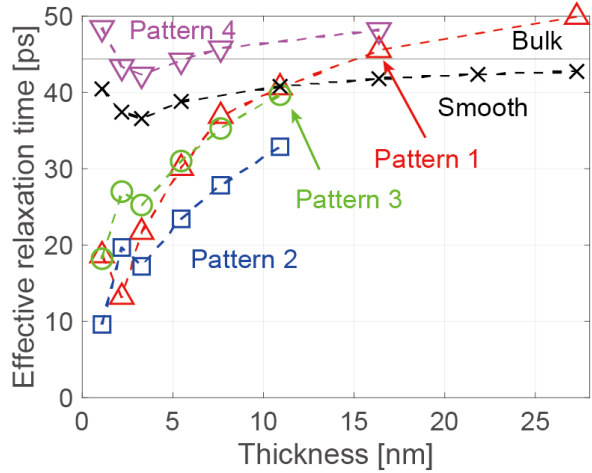

FIG. 10. Thickness dependence of the effective relaxation time at $T$=300 K. Red, blue, green, purple, and black dashed lines denote the $\tau_{\text{eff}}$ for Patterns 1–4 and the smooth thin film, respectively.

### 2. Scattering rate

As mentioned in the previous section, the suppression of the relaxation time due to surface roughness depends significantly on the thickness. To further investigate this, we analyzed the scattering rate $\tau^{-1}$ of the LA mode near the $\Gamma$-point, which contributed significantly to the thermal conductivity (see Supplementary Materials). The contributions to $\tau^{-1}_{qs}$ can be separated into the SPS term, which arises from harmonic properties, and the V3 term from anharmonic properties. We then evaluated the impact of roughness on each. In three-phonon processes, phonon–phonon scattering involves three phonon modes; thus, the target LA mode ($q$, $s$) scatters with two other modes: ($q'$, $s'$) and ($q''$, $s'$). We calculated the frequency-summed results of the weighted SPS and V3 for the interacting modes. The definitions are as follows.

$$\overline{wSPS}(\omega_{\text{in}}) = \sum_{q'\omega'}\sum_{q''\omega''}[-(n_1 + n_2 + 1)\delta(\omega + \omega' + \omega'') + (n_1 + n_2 + 1)\delta(\omega - \omega' - \omega'') - (n_1 - n_2)\delta(\omega - \omega' + \omega'') + (n_1 - n_2)\delta(\omega + \omega' - \omega'')] \times \frac{1}{2}[\delta(\omega_{\text{in}} - \omega') + \delta(\omega_{\text{in}} - \omega'')] \quad (6)$$

$$\overline{V3}(\omega_{\text{in}}) = \left(\frac{\hbar}{2N_q}\right)^3 \sum_{q'\omega'}\sum_{q''\omega''}\left[\frac{1}{\sqrt{\omega_{qs}\omega_{q's'}\omega_{q''s''}}}\sum_{kk'k''}\frac{1}{\sqrt{M_k M_{k'} M_{k''}}}\sum_{\mu\nu\lambda}\Phi_{\mu\nu\lambda}(kk'k'')e_\mu e'_\nu e''_\lambda\right]^2$$
$$\times \frac{1}{2}[\delta(\omega_{\text{in}} - \omega_{q's'}) + \delta(\omega_{\text{in}} - \omega_{q''s''})] \quad (7)$$

Here, $k$, $\Phi(kk'k'')$, and $e$ denote the atom ID, cubic IFCs for the triplet ($k$, $k'$, $k''$), and eigenvector, respectively. $q'$ and $q''$ are any combinations of wavevectors that satisfy the momentum-conservation law with the wavevector $q$ of the analyzed LA mode. $\omega$, $\omega'$, and $\omega''$ represent the phonon frequencies corresponding to branches $s$, $s'$, and $s''$ with the wavevectors $q$, $q'$, and $q''$, respectively.

In Fig. 11(a), the WSPS, which was derived from the harmonic properties, exhibits no significant variation across all frequencies, regardless of the surface roughness. Although the dispersion relations were modulated for Pattern 1 (Fig. 5), the phonon DOS was not modulated (see Supplementary Materials), leading to the same WSPS as for the smooth thin film. However, V3, which was derived from anharmonic properties, exhibited a significantly higher contribution to $\tau^{-1}$ in rough films (Patterns 1–3) compared to the smooth and Pattern 4 films. This indicates that an increase in roughness suppresses the relaxation time through anharmonicity. Additionally, the frequency dependence of the V3 term for the smooth and Pattern 4 films was almost identical. This insensitivity to low roughness ($\eta \simeq 0.01$ nm) was consistent with that of the relaxation time.

Next, we calculated the sum of the cubic IFCs $\Phi(kk'k'')$ for each atom $k$. The results are presented in Fig. 12. As shown in Figs. 12(a)–(c), no modulation of the total cubic IFCs was observed inside the films. However, the cubic IFCs were modulated near the surface. Two scenarios can explain the effect of surface roughness on this IFC modulation: 1) the proximity of the surface reduced the number of neighboring atoms around the surface atoms, reducing the total IFCs of the atoms even if the force constants per bond did not change; 2) the roughness altered the relative coordinates of the surface atoms to the neighboring atoms from those of the single-crystal structure, leading to changes in the IFCs for each bond.

Reductions in the anharmonic IFCs due to Scenario 1 were observed for all thin films, with no apparent effect of surface roughness. However, the increase in anharmonic IFCs attributed to Scenario 2 were observed only for Pattern 1 and were negligible for Pattern 4 and smooth thin film. Therefore, the increases in the anharmonic IFCs due to Scenario 2 were significantly affected by the surface roughness.

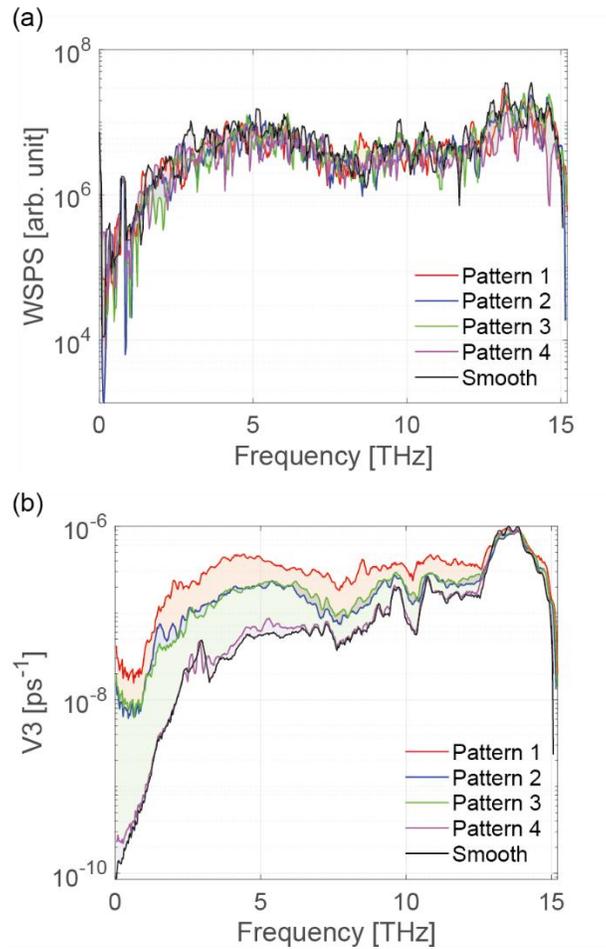

FIG. 11. (a) Frequency-summed WSPS at $T=300$ K and $t=2.19$ nm. (b) Frequency-summed V3 at $T=300$ K and $t=2.19$ nm.

According to Eq. (7), cubic IFCs are the dominant contributors to the V3 term. Therefore, the observed increases in the cubic IFCs led to a corresponding increase in V3, which reduced the relaxation time. As mentioned previously, cubic IFC modulation is observed only in Fig. 12(a), which is consistent with the fact that a significant reduction in the relaxation time occurred only in thin films with high roughness. Additionally, the increases in the cubic IFCs occurred predominantly near the surface, with minimal changes in the internal atoms. Thus, the modulation of cubic IFCs was proportional to the surface-to-volume ratio of the thin film, which explained the thickness dependence of the effective relaxation time.

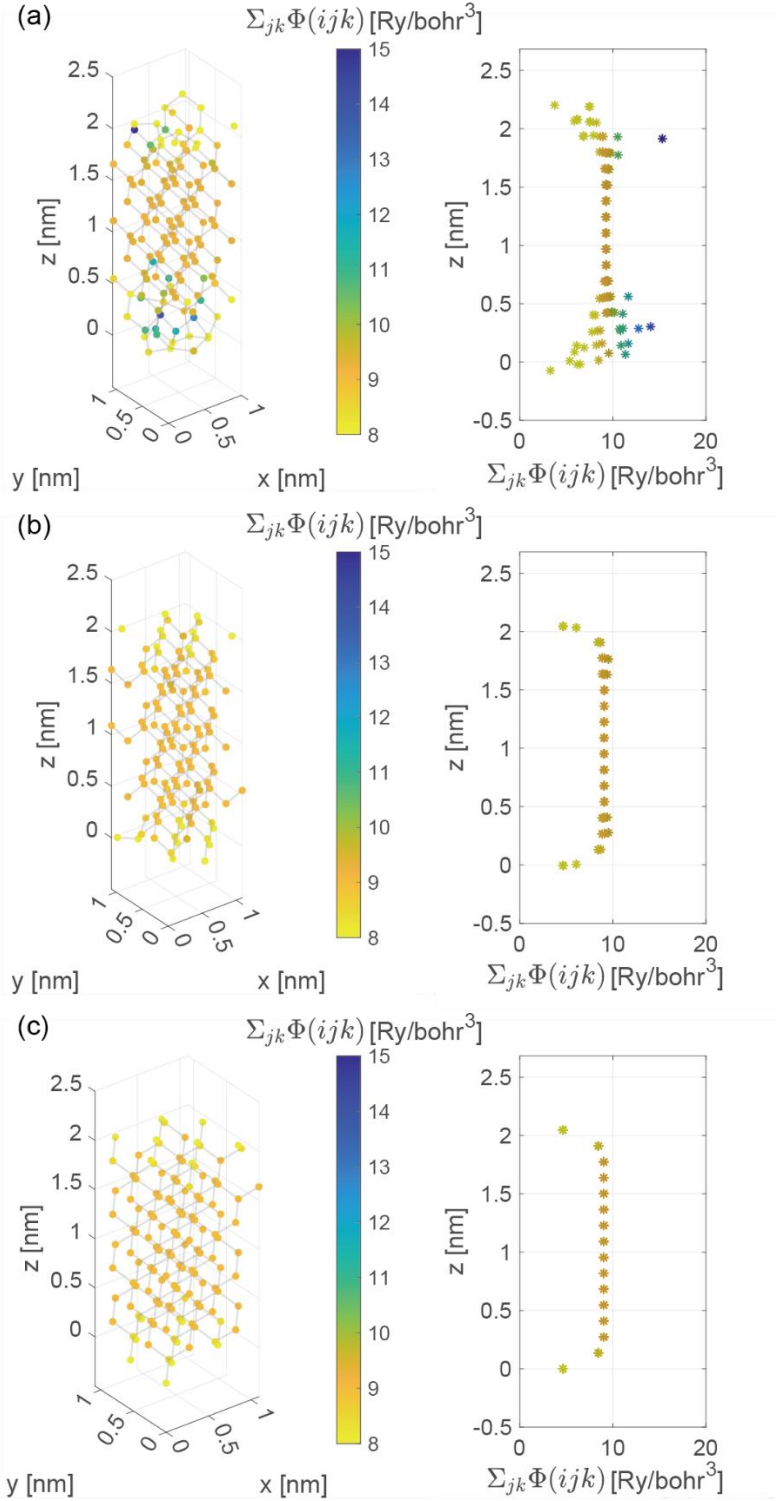

FIG. 12. (a–c) Sum of cubic IFCs for atom $i$ in a thin film with $t=2.19$ nm. The left and right figures show the sum of the cubic IFCs projected onto each atom in the crystal and $z$-coordinate dependence of the sum of anharmonic IFCs for each atom, respectively. (a) Pattern 1; (b) Pattern 4; (c) smooth thin film.

## IV. CONCLUSION

We introduced surface roughness to silicon ultrathin films with thicknesses ranging from approximately 1 to 25 nm and analyzed their thermal conductivities using ALD. The thermal conductivities of thin films with relatively high roughness ($\eta \approx 0.1$ nm), were reduced by up to one order of magnitude compared to those of smooth films and were close to the diffuse-scattering limit and experimental values. Additionally, the introduction of extremely small roughness ($\eta \leq 0.01$ nm) reduced the thermal conductivity by approximately 50% compared to that of smooth films. We analyzed the effects of both the harmonic and anharmonic phonon properties on the thermal-conductivity suppression due to roughness. From a harmonic perspective, the hybridization between surface-localized phonons and heat-conducting phonons reduced the group velocity and thus the thermal conductivity. From an anharmonic perspective, an increase in the anharmonic IFCs of atoms near the surface led to higher scattering rates and thus reduced thermal conductivity. Reduction of group velocity significantly suppressed thermal conductivity, regardless of the film thickness or roughness. Thus, this mechanism predominantly drove the thermal-conductivity reduction in films with $\eta \leq 0.01$ nm. In contrast, anharmonic IFC modulation depends significantly on both film thickness and roughness. Thus, this mechanism dominated the thermal-conductivity reduction in films of $\leq 5$ nm with $\eta \simeq 0.1$ nm.

Because the Ziman-based Fuchs–Sondheimer model is based on the phonon transport properties in the bulk, it cannot account for the group-velocity reduction caused by phonon dispersion modulation. In addition, because Ziman's derivation of the specularity parameter does not include anharmonicity, the theoretical model cannot predict the reduction in relaxation time due to the increased cubic IFCs. Consequently, the theoretical model overestimated the thermal conductivity of the thin films at scales below 20 nm.

## ACKNOWLEDGMENTS
This work was supported by JST-CREST (Grant Numbers JPMJCR21O2 and JPMJCR19I2) and JSPS KAKENHI (Grant Numbers JP22J22155 and 22H04950).